\documentclass[12pt]{article}
\usepackage{amsfonts}
\usepackage{amsmath}

\newcommand{\R} {{\rm I}\!{\rm R}}

\newcommand{\xdif} {
\begin{picture}(12,10)
\put(1,0) {{\LARGE\it x}} \put(2.2,4.4){\line(1,0){8}}
\put(2.2,4.5){\line(1,0){8}} \put(2.2,4.6){\line(1,0){8}}
\end{picture}}

\begin{document}

\title{\bf Submanifolds associated \\ to Toda theories}
\author{\bf E. P. Gueuvoghlanian \\ 
 Instituto de F\' \i sica Te\'orica, 
UNESP, \\ 
Rua Pamplona 145,  01405-900,  S\~ao Paulo, SP, Brazil. \\ 
e-mail:gueuvo@ift.unesp.br}
\date{}

\maketitle

\begin{abstract}
A set of two-dimensional semi-riemannian submanifolds of flat
semi\\-riemannian  manifolds is associated to each Toda theory.
The method and an example are  given to Toda theories associated
to real finite dimensional Lie algebras.
\end{abstract}

\setcounter{equation}{0}
\def\theequation{\thesection. \arabic{equation}}

\section{Introduction}
The subject of this work is concerned with the theory of classical integrability.
Toda theories are two-dimensional relativistic integrable theories. Their field 
equations are expressed in terms of a zero curvature condition. Toda theories associated
to finite dimensional Lie algebras are called conformal Toda theories (see
\cite{Gomes} for a review). In this work we
consider only Toda theories associated to finite dimensional real Lie algebras.

The purpose of this work is to translate the analytical information, that is, the
field equations, into geometrical information. More precisely, we are going to apply
a method by which a system, such that its field equations are given by a set of zero
curvature conditions, associated to a real Lie algebra, in circunstances to be explained, 
can be associated to a set of semi-riemannian
submanifolds of a given semi-riemannian manifold (see \cite{Sym} for a review).

This work is organized as follows: In the sections 2, 3 and 4 we review some basic facts about
Toda theories, semi-riemannian geometry and semi-riemannian submanifolds, respectively. In the
section 5 we review the above cited method. In the section 6 we apply the
method of section 5 to Toda theories
associated to real Lie algebras and give an example. In the appendix a typical calculation
of this work is shown.

\newpage

\section{Toda theories}
In this section we review some basic facts about Toda theories.

Suppose a Lie algebra $G$ can be decomposed as a direct sum of vector spaces
$G= \oplus_i G_i$, $i\in \mathbb{Z}$, where the subspaces $G_i$ are defined by, $Q\in G$,
 $[Q,T]=iT$, \ $ \forall T \in G_i$. The  element $Q$ is called a grading operator. As
a consequence of the Jacobi identity, $ \forall T \in G_i$, $ \forall V \in G_j$, 
$ [T,V] \in G_{i+j}$. Note that if $i \neq j$, then $G_i $ and $G_j$ have only the zero element
$(0 \in G)$ in common.

Toda theories are defined on a flat manifold, the two-dimensional Minkowski space $Mk^2$. In
a natural coordinate system, globally defined, $x^1=x$, $x^2=t$, the metric tensor has constant
components given by $\eta_{11} = -1$, $\eta_{22} = 1 $, $\eta_{12} = \eta_{21} = 0$. We introduce
new coordinates, $z = t+x$, $\bar{z} = t-x$. Then $\partial_t = \partial +\bar{\partial}$ and
$\partial_x = \partial - \bar{\partial}$, where $\partial \equiv \frac{\partial}{\partial z}$
 and $\bar{\partial} \equiv \frac{\partial}{\partial \bar{z} }$.

Toda theories can be obtained as constrained Wess-Zumino-Witten models  \\(WZW). Given a Lie
algebra $G$ and a grading operator $Q$, define $G_+ \equiv \oplus_{i>0} G_i$
and $G_- \equiv  
\oplus_{i<0} G_i$. If the dynamical variable of a WZW model is a group element $g$, use the 
Gauss decomposition \cite{Barut} to write $g= NBM$, where $N = exp (G_-)$, $M = exp(G_+)$ and 
$B = exp (G_0)$. Then the dynamical variable in the Toda theories is the group element B and
the field equations are given by
\[
\bar{\partial} (B^{-1} \partial B) + [ \varepsilon^- , B^{-1} \varepsilon^{+} B ] =0,
\]
where $\varepsilon^+$, $\varepsilon^-$ are constants such that $\varepsilon^+ \in G_1$
and $\varepsilon^- \in G_{-1}$. As
\[
B \{  \bar{\partial} (B^{-1} \partial B) + [ \varepsilon^- , B^{-1}
    \varepsilon^{+} B ] \} B^{-1}= 
\partial(\bar{\partial}B B^{-1})-[\varepsilon^+,B \varepsilon^-B^{-1}],
\]
the field equations can be expressed in a equivalent way by:
\[
\partial(\bar{\partial}B B^{-1})-[\varepsilon^+,B \varepsilon^-B^{-1}]=0.
\]
Note that $Q\in G_0$ because $[Q,Q]=0$ and $G_0$ is a subalgebra as a consequence of the Jacobi
 identity. If the $G_0$ subalgebra is abelian, the model is called abelian.
There is a procedure to get the action of the Toda theories, given the action
of the WZW model \cite{Gomes}.

Let $ A_t(t,x)$ and $A_x(t,x)$ be gauge potentials. Define $A  \equiv A_z = 1/2
(A_t +A_x)$ and $\bar{A}  \equiv A_{\bar{z}} = 1/2 (A_t-A_x)$. The curvature is given by
 $F_{z\bar{z}}= \partial\bar{A} - \bar{\partial} A
  +[A,\bar{A}]$. The Toda theories field equations can
be seen as a zero curvature condition, $F_{z\bar{z}}=0$, where
\begin{eqnarray*}
&& A = B \varepsilon^-  B^{-1} \ \  \hbox{and} \\
&& \bar{A}=- \varepsilon^+ - \bar{\partial} B B^{-1}.
\end{eqnarray*}
Given a group element  $g$ and a gauge transformation
\begin{eqnarray}
&& A^g = g A g^{-1} -\partial g g^{-1} \ \ \hbox{and} \label{eq2.1} \\
&& \bar{A}^g = g \bar{A} g^{-1} -\bar{\partial} g g^{-1}, \label{eq2.2}
\end{eqnarray}
the curvature is transformed to
$ F^{g}_{z\bar{z}} \equiv F_{z\bar{z}}(A^g) = g F_{z\bar{z}} g^{-1}$. Then \\
$ ( F^{g}_{z\bar{z}} = 0 ) \Longleftrightarrow  (F_{z\bar{z}} = 0) $ and the
field equations are invariant by a gauge transformation.

Let $ \lambda \in \R $,
$ \partial_z \lambda = \partial_{ \bar{z} } \lambda
=0$. Define
\begin{eqnarray}
&& a(z,\bar{z}, \lambda) \equiv e^{- \lambda} B \varepsilon^-  B^{-1} \ \
\hbox{and} \label{eq2.3} \\
&&  \bar{a} (z,\bar{z}, \lambda) \equiv - e^{\lambda} \varepsilon^+ -
\bar{\partial} B B^{-1}. \label{eq2.4}
\end{eqnarray} 
The zero curvature condition
\begin{equation}
\partial\bar{a} - \bar{\partial} a + [a,\bar{a}]=0 \label{eq2.5}
\end{equation}
is equivalent to the Toda theories field equations. Note that:
\begin{itemize}
\item{(a)}  
\begin{eqnarray*}
&& a(z,  \bar{z}, \lambda=0)=A(z, \bar{z}) \ \ and \\
&& \bar{a} (z,\bar{z}, \lambda=0) = \bar{A} (z, \bar{z}).
\end{eqnarray*}
\item{(b)}
$a$ and $ \bar{a}$ are obtained applying a gauge transformation to $A$ and
$\bar{A}$ ( equations (\ref{eq2.1}) and (\ref{eq2.2}) ), where
$ g= e^{ \lambda Q }$.
\end{itemize}

\section{Semi-Riemannian geometry}

\setcounter{equation}{0}
\def\theequation{\thesection. \arabic{equation}}

In this section we review some basic facts about semi-riemannian geometry \cite{Neill}. We
consider $C^{\infty}$ differentiable manifolds $M$ (Hausdorff and second
countable) with a complete atlas. $\cal F (M)$ is the set of  $C^{\infty}$ 
differentiable functions $f:M \rightarrow \R$. Given $p \in M$, $T_p (M)$ is
the set of tangent vectors to $M$ at $p$. Given $p \in M$, an open set
$V \subset M$, $p \in V$ and a coordinate system $\varphi : V \rightarrow
\R^m$, $m=\dim (M)$, \\ $\varphi (p)= (x^1 (p),\ldots , x^m (p) )$, then
$\frac{\partial}{\partial x^1} \mid_p$, \ldots, $\frac{\partial}{\partial x^m} \mid_p$
form a basis for the tangent space $T_p (M)$, that is, $v=v^{i}
\frac{\partial}{\partial x^i} \mid_p$,
$\forall v \in T_p(M)$. Let $\phi: M \rightarrow N$ be a $C^{\infty}$ map,
where $M$ and $N$ are manifolds. The differential map of $\phi$ at $p \in M$
is denoted by \\ $d \phi_p : T_p (M) \rightarrow T_{\phi (p)} (N)$. $\xdif (M)$ is
the set of $C^{\infty}$ vector fields on $M$. Given $V, W \in \xdif (M)$, the
Lie
bracket is denoted by $[V,W]$. 

A metric tensor $g$ on $M$ is a symmetric nondegenerate $\cal F (M)$-bilinear
map \\ $g: \xdif(M) \times \xdif(M) \rightarrow \cal F (M)$ of constant
index. The term nondegenerate means that, $\forall p \in M$, given $v \in
T_p(M)$, if $ g(v,w)=0$, $\forall w \in T_p(M)$, then $v=0$. Given a basis to 
$T_p(M)$ and the matrix $(g_{ij}(p))$ which corresponds to $g$ evaluated on
this basis, it follows that $ (g_{ij}(p))$ is invertible. We denote the
inverse by $(g^{ij}(p))$. Given $p \in M$, one can always find an orthonormal
basis for $T_p(M)$. On this basis $(g_{ij}(p))$ is diagonal, $g_{ij}(p)=
\delta_{ij} \epsilon_j$, where $\epsilon_j=\pm 1$, $\forall j \in \{1, \ldots,
m\}$. The signature is $(\epsilon_1, \ldots, \epsilon_m)$ and the index is
defined as the number of negative signs in the signature. The name semi-riemannian
corresponds to a general signature. The riemannian and lorentzian cases
correspond, respectively, to the indexes $0$ and $1$. Let $M$ and $N$ be
semi-riemannian manifolds with metric tensors $g_M$ and $g_N$. An isometry is
a diffeomorphism $\phi: M \rightarrow N$ such that $\phi^* (g_N)=g_M$, where
$\phi^*(g_N)$ is the pullback of $g_N$.

A connection $D$ is a map $D:\xdif(M) \times \xdif(M) \rightarrow \xdif(M)$
such that
\begin{itemize}
\item[(a)] $D_VW$ is $\cal F (M)$-linear in $V$,
\item[(b)] $D_VW$ is $\R$-linear in $W$ and
\item[(c)] $D_V(fW)=(Vf)W +fD_VW$,
\end{itemize}
$\forall f \in \cal F(M)$, $\forall V,W \in \xdif(M)$. $D_VW$ is called the
covariant derivative of $W$ with respect to $V$.

There is a unique connection $D$ such that 
\begin{itemize}
\item[(a)] $[V,W] = D_VW-D_WV$ and
\item[(b)] $Xg(V,W)=g(D_XV,W)+g(V,D_XW)$,
\end{itemize}
$\forall X, V, W \in \xdif(M)$. $D$ is called the Levi-Civita connection of
$M$. These two properties are known as torsion free condition and metric
compatibility.

Let $x^1, \ldots, x^m$ be a coordinate system on an open set $V \subset M$. The
Christoffel symbols are defined by $D_{\partial_i}\partial_j = \Gamma^k_{ij} \partial_k$,
$(1 \leq i,j \leq m)$. In the case of a Levi-Civita connection $\Gamma^k_{ij} =
\Gamma^k_{ji}$ and
\begin{equation}
\Gamma^k_{ij} = \frac{1}{2} g^{kr} \left( \frac{\partial g_{jr}} {\partial x^i} 
+ \frac{\partial g_{ir}} {\partial x^j} - \frac{\partial g_{ij}} {\partial x^r}
\right), \label{eq3.1}  
\end{equation}
$\forall i,j,k \in \{1, \ldots m \}$.

The Riemann tensor $R : [ \xdif (M) ]^3 \rightarrow \xdif(M)$ is defined by
\\ $ R(X,Y)Z= D_{[X,Y]}Z-[D_X,D_Y]Z$, $\forall X, Y, Z \in \xdif(M)$ and is $\cal F
(M)$-linear in $X, Y$ and $Z$. Given a coordinate system, $ R(\partial_k,\partial_l)\partial_j
=R^i_{jkl} \partial_i$, where
\begin{equation}
R^i_{jkl} \equiv \frac{\partial \Gamma^i_{kj}}{\partial x^l} - \frac{\partial \Gamma^i_{lj}}{\partial x^k}
        +  \Gamma^i_{lr} \Gamma^r_{kj} - \Gamma^i_{kr} \Gamma^r_{lj} \label{eq3.2}   
\end{equation}
and we define $R_{ijkl} \equiv g_{ir} R^r_{jkl}$, $\forall i,j,k,l \in \{1,
\ldots, m \}$. Different definitions of the Riemann tensor (concerning its
sign) and its components $R^i_{jkl}$ can be found.

Given $p \in M$, a two-dimensional subspace $P$ of $T_p(M)$ is called a tangent
plane to $M$ at $p$. If $g$ restricted to $P$ is nondegenerate, $P$ is said to
be a nondegenerate tangent plane to $M$ at $p$. In this case take a basis
$v,w$ for $P$ and define
\[
K(P) \equiv \frac{g(R(v,w)v,w)}{Q(v,w)},
\]
where $Q(v,w) \equiv g(v,v)g(w,w)-[g(v,w)]^2$. Then $K(P)$ is independent of
the choice $v,w$ for $P$ and is called the sectional curvature of $P$. One can
verify that $K(P)=0$ for every nondegenerate plane $P$ in $T_p(M)$ if and only
if the Riemann tensor satisfies $R(p)=0$, that is, $R(x,y)z=0$, $\forall x,y,z
\in T_p(M)$. $M$ is said to be flat if $R(p)=0$, $\forall p \in M$. $M$ is
said to have constant curvature if $K(P)$ is constant for every $P$ in
$T_p(M)$ and for every $p \in M$.

The Ricci tensor is a $\cal F(M)$-bilinear symmetric map $Ric :[ \xdif (M)]^2
\rightarrow \cal F(M)$ whose components are given by $R_{ij}=R^k_{ijk}$,
$\forall i,j \in \{1, \ldots, m\}$. $M$ is said to be Ricci flat if $Ric
(p)=0$, $\forall p \in M$. The scalar curvature is given by $S=g^{ij} R_{ij}$.

Let $\R^m$ be the $C^\infty$ manifold with its usual differentiable
structure. A natural coordinate system on $\R^m$ is one (globally defined)
that associates to \\ $(x^1, \ldots, x^m) \in \R^m$ the coordinates $(x^1,
\ldots, x^m)$. We define a metric tensor on  $\R^m$ in such a way that in a
natural coordinate system the components of $g$ are given by
$g_{ij}(p)=\delta_{ij} \epsilon_j$, $\forall p \in \R^m$, $\forall i,j \in
\{1, \ldots, m\}$, where $\epsilon_j=-1$ for $1 \leq j \leq \nu$ and
  $\epsilon_j=1$, for $\nu +1 \leq j \leq m$. The number $\nu$ is the
  index. This manifold with a Levi-Civita connection is denoted by
  $\R^m_{\nu}$. In natural coordinates $\Gamma^k_{ij}(p)=0$, $\forall p \in
  \R^m_{\nu}$, \\ $\forall i,j,k \in \{1, \ldots, m\}$.  The Riemann tensor
  satisfies $R(p)=0$, $\forall p \in \R^m_{\nu}$. These flat manifolds are
  called semi-euclidean spaces. $\R^m_0$ is the $m$-dimensional Euclidean space
  $E^m$ and, if $m \geq 2$, $\R^m_1$ is the $m$-dimensional Minkowski space $Mk^m$.

Let $M$ be a two-dimensional manifold. Then $T_p(M)$ is the only tangent
plane at $p \in M$. The sectional curvature in this case is denoted by $K$ and
is called the gaussian curvature. Then $R(X,Y)Z=K[g(Z,X)Y-g(Z,Y)X]$, $\forall
X, Y, Z \in \xdif(M)$, $Ric = Kg$, $S=2K$ and $K=-R_{1212}$.

\section{Semi-Riemannian submanifolds}

\setcounter{equation}{0}
\def\theequation{\thesection. \arabic{equation}}

In this section we review some basic facts about semi-riemannian
submanifolds \cite{Neill}. At first place, submanifolds, immersions and imbeddings are
defined. Then it is shown that, given an immersion, there is always a
submanifold locally defined. The next concept introduced is that of
semi-riemannian submanifolds. By last, the Gauss-Weingarten equations and the
Gauss-Codazzi-Ricci equations are discussed.

Let $\bar{M}$ be a topological space and $M \subset \bar{M}$. The induced
topology is the one such that a subset $V$ of $M$ is open if and only if there
is an open set $\bar{V}$ of $\bar{M}$ such that $\bar{V} \cap M = V$. $M$ is a
topological subspace of $\bar{M}$ if it has the induced toplogy. A manifold
$M$ is a submanifold of a manifold $\bar{M}$ provided:
\begin{itemize}
\item[(a)] $M$ is a topological subspace of $\bar{M}$.
\item[(b)] The inclusion map $j: M \subset \bar{M}$ is $C^{\infty}$ and at
  each point $p \in M$ its differential map $dj$ is injective.
\end{itemize}

 Let $\bar{M}$
  and $N$ be $C^{\infty}$ manifolds. An immersion $\phi : N \rightarrow
  \bar{M}$ is a  $C^{\infty}$ map such that $d \phi_p$ is injective, $\forall
  p \in N$. An imbedding of a manifold $N$ into $\bar{M}$ is an injective
  immersion $\phi: N \rightarrow \bar{M}$ such that the induced map $N
  \rightarrow \phi (N)$ is a homeomorphism onto the subspace $\phi (N)$ of
  $\bar{M}$, where $\phi (N) \subset \bar{M}$ has the induced topology. An
  open set $V$ of a manifold $N$ is a manifold, considering the restriction of
  the complete atlas of $N$ to $V$. Let $\phi: N \rightarrow \bar{M}$ be an
  immersion. $\forall p \in N $, there is an open set $V \subseteq N$, $p \in V$,
  such that $\phi: V \rightarrow \bar{M}$ is an imbedding \cite{CarmoRie}. If $\phi: V
  \rightarrow \bar{M}$ is an imbedding, make its image $\phi(V)$ a manifold so
  that the induced map $\hat{\phi} : V \rightarrow \phi(V) $ is a
  diffeomorphism. Then $\phi (V)$ is a topological subspace of $\bar{M}$ and
  the inclusion map $j: \phi(V) \subset \bar{M}$ is $\phi \circ
  [\hat{\phi}]^{-1}$, which by the chain rule is an immersion. Thus $\phi(V)$
  is a submanifold of $\bar{M}$ \cite{Neill}.

Let $M$ be a submanifold of the semi-riemannian manifold $\bar{M}$ with the
metric tensor $\bar{g}$. If the pullback $j^* (\bar{g})$ is a metric tensor on
$M$, then $M$ is called a semi-riemannian submanifold of $\bar{M}$. In the
particular case of a riemannian manifold $\bar{M}$, the previous condition is
always true.

Let us explain better this last topic. If $p \in M$ and $g$ is the metric
tensor on the semi-riemannian submanifold $M$, then $g(v,w)= \bar{g}
(dj_p(v),dj_p(w)) \equiv \bar{g} (v,w)$, $\forall v,w \in T_pM$, where we have
simplified our notation, as it is usually done, by the identification of
$T_pM$ and $dj_p(T_p(M))$, which turns out to be equivalent to say $T_p(M)
\subset T_p( \bar{M} )$. The previous condition has to be verified in order to
define a semi-riemannian submanifold, because the restriction of $\bar{g}$ to
$T_p(M) \times T_p(M)$, $\forall p \in M$, in some cases does not define a
metric tensor on $M$. The metric tensor $g$ is called the first fundamental
form of the submanifold $M$. There is another usual simplification in the
notation. We write $\langle , \rangle$ for the scalar product defined by
$\bar{g}$ and understand that, when it is restricted to $T_p(M) \times
T_p(M)$, $\forall p \in M$, it is the scalar product defined by $g$.

Suppose $M$ is a semi-riemannian submanifold of $\bar{M}$. Then define,
$\forall p \in M$,
\[
T_p(M)^{\perp}= \{v \in T_p(\bar{M}) \mid \bar{g} (v,w)=0, \forall w \in
T_p(M) \}.
\]
As $\bar{g}$ restricted to $T_p(M) \times T_p(M)$ is nondegenerate,
$T_p(\bar{M})= T_p(M) \oplus T_p(M)^{\perp}$ and $\bar{g}$ restricted to $T_p(M)^{\perp}
\times T_p(M)^{\perp}$ is also nondegenerate. The subspaces in the direct sum
are called tangent and normal subspaces. The projections are denoted by $ Tan:
T_p(\bar{M}) \rightarrow T_p(M)$ and $Nor: T_p(\bar{M}) \rightarrow
T_p(M)^{\perp}$, $\forall p \in M$.

We denote by $ \bar{\xdif} (M) $ the set of $C^{\infty}$ vector fields defined
on $M$ such that if $X \in \bar{\xdif} (M) $, then $X(p) \in T_p(\bar{M})$,
$\forall p \in M$. Similarly, the two sets of  $C^{\infty}$ vector fields
defined on $M$  denoted by $\xdif (M) $ and $ \xdif (M) ^{\perp}$ are such that, if
$ Y \in  \xdif (M) $ and $Z \in  \xdif (M) ^{\perp}$, then $Y(p) \in
T_p(M)$ and $Z(p) \in T_p(M)^{\perp}$, $\forall p \in M$.

We write $\bar{D} : \xdif(\bar{M}) \times \xdif(\bar{M}) \rightarrow
\xdif(\bar{M})$ for the Levi-Civita connection on $\bar{M}$, where
$\xdif(\bar{M})$ is the set of $C^{\infty}$ vector fields on $\bar{M}$. There
is a natural defined induced connection  $\bar{D} : \xdif(M) \times \bar{\xdif}(M) \rightarrow
\bar{\xdif}(M)$, obtained from the Levi-Civita connection $\bar{D}$ on
$\bar{M}$ by taking appropriate $C^{\infty}$ local extensions of the vector
fields in the sets $\xdif (M)$ and $ \bar{\xdif} (M)$ to $\xdif(\bar{M})$. The
induced connection has analogous properties to the Levi-Civita connection
$\bar{D}$ on $\bar{M}$.

The Gauss-Weingarten equations are:
\begin{eqnarray}
&&\bar{D}_VW= Tan(\bar{D}_VW)+Nor(\bar{D}_VW) \ \ \hbox{and} \label{eq4.1} \\
&&\bar{D}_VZ= Tan(\bar{D}_VZ)+Nor(\bar{D}_VZ), \label{eq4.2}
\end{eqnarray}
$\forall V,W \in \xdif(M)$, $Z \in \xdif(M)^{\perp}$, with the following identifications:
\begin{itemize}
\item[(a)] $ Tan(\bar{D}_VW)= D_VW$, where $D: \xdif (M) \times \xdif (M)
  \rightarrow \xdif (M)$ is the Levi-Civita connection on $M$.
\item[(b)] $Nor(\bar{D}_VW)= \Pi (V,W)$, where $\Pi: \xdif (M) \times \xdif (M)
  \rightarrow \xdif(M)^{\perp}$ is $\cal F (M)$-bilinear and symmetric. $\Pi$
  is denominated the second fundamental form of the submanifold $M$.
\item[(c)] $Tan(\bar{D}_VZ)= \tilde{\Pi} (V,Z)$, where $\tilde{\Pi} : \xdif (M) \times
  \xdif (M)^{\perp} \rightarrow \xdif(M)$ is $\cal F (M)$-bilinear. $ \tilde{\Pi}$
  is given in terms of $\Pi$, because $\langle \tilde{\Pi} (V,Z),W \rangle = -
  \langle \Pi (V,W),Z \rangle$, $\forall V,V \in \xdif(M)$, $Z \in
  \xdif(M)^{\perp}$.
\item[(d)] $Nor(\bar{D}_VZ)= D^{\perp}_V Z$, where $D^{\perp}_V Z$ is called
  the normal covariant derivative of $Z$ with respect to $V$ and $D^{\perp}$
  is called the normal connection on $M$.
\end{itemize}

One can verify that
\begin{itemize}
\item[(a)] $D^{\perp}_VZ$ is $\cal F(M)$-linear in $V$ and $\R$-linear in $Z$,
\item[(b)] $D^{\perp}_V(fZ)= f D^{\perp}_VZ +(Vf)Z$ and
\item[(c)] $ V \langle Y,Z \rangle= \langle D^{\perp}_VY,Z \rangle + 
             \langle Y, D^{\perp}_VZ \rangle$ ,
\end{itemize}
$\forall V \in \xdif (M)$, $Y,Z \in \xdif(M) ^{\perp}$ and $f\in \cal F (M)$. The
normal connection plays on the normal bundle $NM$ the role played by the
Levi-Civita connection $D$ on the tangent bundle $TM$.

The maps $g$, $\Pi$ and $D^{\perp}$ are called the fundamental forms of the
submanifold $M$. As a consequence of the Gauss-Weingarten equations \cite{Neill}, we have:
\begin{itemize}
\item[(a)] The Gauss equation:
\[
\langle R(V,W)X,Y \rangle = \langle \bar{R}(V,W)X,Y \rangle  +\langle
\Pi(V,X), \Pi (W,Y) \rangle - \langle \Pi(V,Y), \Pi (W,X) \rangle,
\]
$\forall V$, $W$, $X$, $Y \in \xdif(M)$, where $\bar{R}$, $R$ are the Riemann
tensors on $\bar{M}$ and $M$, respectively.

\item[(b)] The Codazzi equation:
\[
Nor \bar{R} (V,W) X = - (\bigtriangledown_V \Pi) (W,X) + (\bigtriangledown_W
\Pi) (V,X),
\]
where 
\[ 
(\bigtriangledown_V \Pi) (X,Y)= D^{\perp}_V ( \Pi (X,Y)) - \Pi( D_VX,Y) - \Pi
(X, D_VY),
\]

$\forall V,W,X,Y \in \xdif(M)$.

\item[(c)] The Ricci equation:
\[
\langle R^{\perp}(V,W)X,Y \rangle = \langle \bar{R}(V,W)X,Y \rangle  +\langle
\tilde{\Pi}(V,X), \tilde{\Pi} (W,Y) \rangle - \langle \tilde{\Pi}(V,Y), \tilde{\Pi} (W,X) \rangle
\]

$\forall V,W \in \xdif (M)$, $X,Y \in \xdif(M)^{\perp}$, where
\\ $R^{\perp} : \xdif(M) \times \xdif(M) \times \xdif(M)^{\perp}  \rightarrow
\xdif(M)^{\perp} $ 
is given by
\\ $ R^{\perp}(V,W)X = D^{\perp}_{[V,W]} X - [D^{\perp}_V, D^{\perp}_W ]X$ and is
$\cal F(M)$-multilinear.
\end{itemize}

One can get a component representation of the Gauss-Weingarten (GW) equations
and the Gauss-Codazzi-Ricci (GCR) equations. Let $\dim(\bar{M}) =
\bar{m}$, \\ $\dim(M)=m$ and adopt the following convention about indexes: latin
lower case letters $(a, b, \ldots )$ take values on the set $\{ 1, \ldots,
\bar{m} \}$, latin upper case letters $\{ A, B, \ldots \} $ take values on the
set $\{ 1, \ldots, \bar{m} -m \}$ and greek letters $ \{ \alpha, \beta,
\ldots \}$ take values on the set $\{ 1, \ldots, m \}$. The summation
convention over indexes appearing twice, once as a superscript and once as a
subscript, is employed.

Let $\bar{V}$ be a coordinate neighborhood of $p \in \bar{M}$ and $V$ a
coordinate neighborhood of $p \in M$, $p \in V \subset \bar{V}$, such that
there is a set of $(\bar{m} - m)$ locally defined $C^{\infty}$ vector fields
$\{N_A\}$, $N_A(q) \in T_q(M)^{\perp}$, $\forall q \in V$, $\langle N_A, N_B
\rangle = \eta_{AB}$, where $\eta_{AB}(q)= \delta_{AB} \epsilon_B^{\perp}$ and 
$\epsilon_B^{\perp}= \pm 1$. The index corresponding to the signature 
$( \epsilon_1^{\perp}, \ldots, \epsilon _{\bar{m}-m}^{\perp})$ is
denoted by $\nu^{\perp}$. We denote the inverse of the matrix $(\eta_{AB})$ 
by $(\eta^{AB})$. Similarly, $(\bar{g}^{ij})$ and $(g^{\mu\nu})$ are the
inverses of  $(\bar{g}_{ij})$ and $(g_{\mu\nu})$. The normal frame field
$ \{ N_A \} $ is typically employed in order to construct bundle charts on the
normal bundle $NM$ \cite{Neill}. We denote by $\{ \bar{\Gamma}^k_{ij} \}$ and 
$\{ {\Gamma}^{\gamma}_{\alpha\beta} \}$ the Christoffel symbols of
$\bar{M}$ and $M$ respectively.

Given local coordinates $(y^1, \ldots, y^{\bar{m}})$ on $\bar{V}$, $(z^1,
\ldots, z^m)$ on $V$, then $\partial_{\alpha}= y^k_{,\alpha} \partial_k$, 
where $\partial_{\alpha} \equiv \frac{\partial}{\partial z^{\alpha}}$, 
$\partial_k \equiv \frac{\partial}{\partial y^k}$ and $y^k_{,\alpha} \equiv
\frac{\partial y^k}{\partial z^{\alpha}}$. Define $\Pi (\partial_{\alpha},
\partial_{\beta}) \equiv \eta^{AB} b_{B\alpha\beta} N_A$ and
$D^{\perp}_{\partial_{\alpha}} N_A \equiv \mu^B_{ \, \, \, \, A \alpha} N_B$. Then
$\tilde{\Pi} (\partial_{\alpha}, N_A) = - b_{A\alpha\beta} g^{\beta\gamma}
\partial_{\gamma}$. The set of functions  \\ $ \{ \{ g_{\alpha\beta} \}, \{
b_{A\alpha\beta} \}, \{ \mu^A_{ \, \, \, \, B\alpha} \} \}$  gives the component representation
of the fundamental \\ forms $(g, \Pi, D^{\perp})$. Note that $b_{B\alpha\beta}=b_{B\beta\alpha}$.

Taking $V= \partial_{\alpha}$, $W=\partial_{\beta}$ and $Z=N_A$ in the
equations (\ref{eq4.1}) and (\ref{eq4.2}), the GW equations can be written as:
\begin{eqnarray}
&& y^k_{,\alpha\beta} = \Gamma^{\gamma}_{\alpha\beta}y^k_{,\gamma} 
-\bar{\Gamma}^k_{pq} y^p_{,\alpha} y^q_{,\beta} + \eta^{AB} b_{A\alpha\beta}
N^k_B \ \ \hbox{and} \label{eq4.3} \\
&& N^i_{A,\alpha} = - \bar{\Gamma}^i_{pq} y^p_{,\alpha} N^q_A + 
\mu^C_{ \, \, \, \, A \alpha} N^i_C - b_{A\alpha\gamma} g^{\gamma\rho} y^i_{,\rho} \ ,
  \label{eq4.4}
\end{eqnarray}
where $N_A \equiv N_A^i \partial_i$ and $y^k_{, \alpha\beta} \equiv
\partial_{\alpha}\partial_{\beta}y^k$.

We define $r^C_{ \, \, \, \, A\alpha\beta}$ by $
R^{\perp}(\partial_{\alpha},\partial_{\beta})N_A \equiv r^C_{ \, \, \, \, A\alpha\beta}
N_C$. Thus 
\[
r^C_{ \, \, \, \, A\alpha\beta}= \partial_{\beta} \mu^C_{ \, \, \, \,  A \alpha} + 
\mu^B_{ \, \, \, \, A \alpha}  \mu^C_{ \, \, \, \, B \beta}
- \partial_{\alpha} \mu^C_{ \, \, \, \, A \beta}
- \mu^B_{ \, \, \, \, A \beta}  \mu^C_{ \, \, \, \, B \alpha}.
\]
The GCR equations can be written as:
\begin{itemize}
\item[(a)] The Gauss equation:
\[
R_{\delta\gamma\alpha\beta} = y^i_{,\alpha}  y^j_{,\beta}  y^k_{,\gamma}
y^l_{,\delta} \bar{R}_{lkij}
+ \eta^{CD} ( b_{C\alpha\gamma} b_{D\beta\delta} -  b_{C\alpha\delta}
b_{D\beta\gamma} ).
\] 
\item[(b)] The Codazzi equation:
\[
\bar{R}_{mkij} y^i_{,\alpha}y^j_{,\beta}y^k_{,\gamma} N^m_D=
b_{B\beta\gamma} \mu^B_{ \, \, \, \, D\alpha} - b_{B\alpha\gamma} \mu^B_{ \,
  \, \, \, \, D\beta}
-b_{D\beta\gamma;\alpha} + b_{D\alpha\gamma;\beta} ,
\]
where
\[
b_{D\beta\gamma;\alpha} \equiv \partial_{\alpha} b_{D\beta\gamma}
- \Gamma^\theta_{\alpha\gamma} b_{D\beta\theta}
-  \Gamma^\theta_{\alpha\beta} b_{D\theta\gamma}.
\]
\item[(c)]  The Ricci equation:
\[
\eta_{CB} r^C_{ \, \, \, \, A\alpha\beta} = y^i_{,\alpha} y^j_{,\beta} N^k_A N^l_B
\bar{R}_{lkij} +b_{A\alpha\gamma} b_{B\beta\theta} g^{\theta\gamma}
-b_{B\alpha\gamma}b_{A\beta\theta} g^{\theta\gamma}.
\]
\end{itemize}

We raise and lower latin uppercase / latin lowercase / greek indexes with
\\ $(\eta^{AB}), (\eta_{AB})$ / $(g^{ij}), (g_{ij}) $ / $( \bar{g}^{\alpha\beta})
, ( \bar{g}_{\alpha\beta}) $. These six matrices are symmetric ones.

As $ (\eta_{AC})$ is constant in $V$, 
\begin{eqnarray*}
0 &=& \partial_{\alpha} \langle N_A,N_C \rangle = \langle D^{\perp}_{\alpha}
N_A, N_C \rangle + \langle N_A , D^{\perp}_{\alpha} N_C \rangle \\
  &=& \mu^B_{ \, \, \, \, A\alpha} \eta_{BC} +  \mu^B_{ \, \, \, \, C\alpha} \eta_{AB} 
\equiv \mu_{CA\alpha}+ \mu_{AC\alpha}.
\end{eqnarray*}
Then $ \mu_{CA\alpha}= -\mu_{AC\alpha}$. Thus
$\mu_{11\alpha}=\mu_{22\alpha}=\ldots=
\mu_{ (\bar{m}-m) (\bar{m}-m)\alpha } =0$.

Let $M$ be a hypersurface, that is, $\bar{m}= m+1$. In this case, latin
uppercase indexes take just one value (1). Then, as $ \mu^A_{ \, \, \, \, C\alpha} =
\eta^{AB} \mu_{BC\alpha}$, $\mu^1_{ \, \, \, \, 1\alpha}=0$ and $ r^1_{ \, \,
  \, \, 1\alpha\beta}=0$. $\bar{R}$ satisfies $\bar{R}_{lkij}=-\bar{R}_{klij}$. Thus
both sides of the Ricci equation are equal to zero and the equation is
trivially satisfied in this case.

The mean curvature vector field $\vec{\cal H} $ is defined by:
\begin{equation}
\vec{{\cal H}} (p) =
\frac{1}{m} \sum^{m}_{\alpha=1} \epsilon_{\alpha} \Pi (e_{\alpha},
e_{\alpha}), \label{eq4.5}
\end{equation}
where $ \{ e_{\alpha} \} $ is any orthonormal basis for $T_p(M)$, $g(e_{\alpha}, 
e_{\beta})=\delta_{\alpha\beta} \epsilon_{\beta}$ and $\epsilon_{\beta}= \pm 1$.

In the classical theory of surfaces in the three-dimensional euclidean space
$E^3$, the gaussian curvature $K$ and the mean curvature ${\cal H}  $ are defined
in terms of the differential of the normal map of Gauss in such a way that $
{\cal H} (p) =\frac{k_1+k_2}{2}$ and $K(p)=k_1k_2$, where $k_1$, $k_2$ are the
principal curvatures of the surface $S$ at $p \in S$. The result obtained by
Gauss that $K$ is invariant by local isometries is known as the Egregium
Theorem \cite{CarmoDif}.

One can ask if given functions $ \{ \{ g_{\alpha\beta} \}, \{ b_{A\alpha\beta}
  \}, \{ \mu^A_{ \, \, \, \, B\alpha} \} \}$ satisfying the GCR equations is there a
  submanifold such that these functions are the corresponding components of
  the fundamental forms. The answer to this question in the classical theory
  of surfaces in $E^3$ is known as the Bonnet theorem and, basically, it says
  that locally, up to rigid motions, there is one such surface. See the
  reference \cite{CarmoDif} for the precise statement of this theorem. About the generalized
  Bonnet theorem on  manifolds \\ $\bar{M}$ with constant sectional curvature,
  also called the fundamental theorem of submanifolds, see \cite{Spivak}.

\section{Submanifolds defined by zero curvature \\ conditions}

\setcounter{equation}{0}
\def\theequation{\thesection. \arabic{equation}}

In this section we review a method by which a system, such that its field
equations are given by a set of zero cuvature conditions, associated to a real
Lie algebra, in circunstantes to be explained, can be associated to a set of
semi-riemannian submanifolds of a given semi-riemannian manifold (see
\cite{Sym} for a review).

The first step is to associate to a given real Lie algebra $G$ a manifold with
a metric tensor and a corresponding Levi-Civita connection. One of the basic
ideas is that all $\bar{m}$-dimensional vector spaces $V$ over $\R$ are
isomorphic to the vector space $\R^m$ by the choice of a basis in
$V$. Corresponding to a $\bar{m}$-dimensional Lie algebra $G$ there is a 
$\bar{m}$-dimensional vector space. We can associate to $\R^m$ a manifold with
its usual differentiable structure. In this way we can associate to a
$\bar{m}$-dimensional Lie algebra $G$ a differentiable manifold which is
diffeomorphic to the differentiable manifold $\R^{\bar{m}}$.

The Killing form \cite{Cornwell} on a real Lie algebra $G$ is a symmetric $\R$-bilinear map 
\\ $k: G \times G \rightarrow \R$. This map is nondegenerate if and only if the
Lie algebra is semisimple. From now on we will suppose the Lie algebra is
semisimple. As $k$ is nondegenerate it is possible to find an orthonormal
basis $ \{ T_i \} $, $T_i \in G$, that is, $k(T_i,T_j) = \delta_{ij}
\bar{\epsilon}_j$, $\bar{\epsilon}_j = \pm 1$. We denote by $\bar{\nu}$ the
index of $k$, that is, the number of negative signs in the signature $(
\bar{\epsilon}_1, \ldots, \bar{\epsilon}_{\bar{m}} )$. The Killing form is
invariant by automorphisms and satisfy
\begin{equation}
k([T_i,T_j],T_r)=k(T_i,[T_j,T_r]). \label{eq5.1}
\end{equation}

Let us denote for a moment the manifold associated to the Lie algebra $G$ by
$\R^{\bar{m}}_G$. There is a diffeomorphism relating $\R^{\bar{m}}$ and 
$\R^{\bar{m}}_G$. Then there is an isomorphism given by the differential of
this diffeomorphism relating the tangent spaces of
these two manifolds at corresponding points. In this way a natural coordinate
system on  $\R^{\bar{m}}$ induces a globally defined coordinate system 
$ \{ y^i \} $ on  $\R^{\bar{m}}_G$, which we also call natural, and
corresponding coordinate fields $ \{ \partial / \partial y^i \} $.

We introduce a set of bijective (injective and surjective) linear maps,
$\forall p \in \R^{\bar{m}}_G$, $L_p:T_p(\R^{\bar{m}}_G) \rightarrow G$, which are defined in
terms of the basis $ \{ \partial / \partial y^i \}$ as \\ $L_p ( \partial /
\partial y^i (p)) \equiv T_i$,
where $ \{ T_i \}$ is the orthonormal basis defined by the Killing form. That
is, each one of the $L_p$ in the set $ \{ L_p \}$ is an isomorphism. We denote
the inverse by  $L_p^{-1}: G \rightarrow T_p(\R^{\bar{m}}_G)$, which is also
bijective and linear. Thus $L_p^{-1} (T_i) = \partial / \partial y^i (p)$.

Let $c \in \R $, $c \neq 0$. Then we introduce a ${\cal
F}(\R^{\bar{m}}_G)$-bilinear map \\ $\bar{g}: \xdif (\R^{\bar{m}}_G) \times \xdif
(\R^{\bar{m}}_G) \rightarrow  {\cal F}    (\R^{\bar{m}}_G)$ which is defined in terms
of the basis $ \{ \partial_i \} $ as
\begin{equation}
\bar{g} ( \partial_i(p), \partial_j (p) ) \equiv c k(L_p (\partial_i(p) ), L_p
(\partial_j (p)))= c k(T_i,T_j)= c \delta_{ij} \bar{\epsilon}_j, \label{eq5.2}
\end{equation}
$ \forall p \in \R^{\bar{m}}_G$. One can see the map $\bar{g}$ is a metric
tensor on $\R^{\bar{m}}_G$. Note that 
\[
\bar{g} \left( \frac{ \partial_i (p) } { \sqrt { \mid c \mid } },  
          \frac{ \partial_j (p) } { \sqrt{ \mid c \mid } } \right)
        = \frac {c}{ \mid c \mid } \delta_{ij} \bar{\epsilon}_j.
\]
That is, $ \{  \partial _i(p) /   \sqrt{ \mid c \mid }   \} $ is an orthonormal
basis at $T_p(\R^{\bar{m}}_G)$. 

Define $\bar{\nu} (c)$ by $\bar{\nu} (c) \equiv \bar{\nu}$ if $c > 0$ and  
$\bar{\nu} (c) \equiv \bar{m} - \bar{\nu}$ if $c < 0$. Then 
$\R^{\bar{m}}_G$ with the metric tensor $\bar{g}$ and a corresponding
Levi-Civita connection is isometric to the semi-euclidean space
$\R^{\bar{m}}_{\bar{\nu} (c)}$.
To simplify our notation we will just denote this manifold by $\R^{\bar{m}}_{\bar{\nu} (c)}$.

We consider field theories such that their field equations are given in terms
of a set of semisimple real Lie algebra valued gauge potentials  $a_1 (z^1,
\ldots, z^m, \lambda)$,$\ldots$, $a_m(z^1, \ldots, z^m, \lambda) $ defined on
$\R^m_{\nu} \times \R$ satisfying a set of zero curvature conditions
\begin{equation}
\partial_{\alpha} a_{\beta} - \partial_{\beta} a_{\alpha}
+[a_{\alpha},a_{\beta}]=0, \label{eq5.3}
\end{equation}
$\forall \alpha, \beta \in \{ 1, \ldots, m \}$.

Let $U$ be defined by: 
\begin{equation}
( \partial_{\mu} + a_{\mu} ) U=0, \label{eq5.4}
\end{equation}
$\forall \mu \in \{ 1,\ldots, m \} $ .

We now introduce an expression which is supposed to parametrize a submanifold
$M$. After we have derived expressions for the corresponding components of the
fundamental forms, we are going to indicate the conditions that must be
satisfied in order to have a consistent construction and a well defined
semi-riemannian \\ submanifold. The maps $L_p$ and $L_p^{-1}$ are implicitly
employed in order to identify $T_p(\R^{\bar{m}}_{\bar{\nu}(c)} )$ and $G$,
$\forall p \in \R^{\bar{m}}_{\bar{\nu}(c)}$.

Let the position vector corresponding to points in the submanifold $M$ be given by
\begin{equation}
r=y^1(z^1,\ldots,z^m, \lambda) \partial_1 +\cdots+
  y^{\bar{m}}(z^1,\ldots,z^m, \lambda) \partial_{\bar{m}} \equiv
  U^{-1}U_{,\lambda} \ \ , \label{eq5.5}
\end{equation}
where $ U_{,\lambda} \equiv  \frac{\partial U}{\partial \lambda}$, $ \{ z^{\alpha}
\}$ are local coordinates on $M$ and $ \{ y^i \} $ are natural coordinates on 
$\R^{\bar{m}}_{\bar{\nu}(c)} $.

Then, using (\ref{eq5.4}), 
\begin{eqnarray}
r_{,\mu} = -U^{-1} a_{\mu,\lambda} U \ \ \hbox{and} \label{eq5.6} \\
r_{,\mu\nu} = U^{-1} ( [a_{\mu,\lambda},a_{\nu}]-a_{\mu,\nu\lambda})U, 
\end{eqnarray}
where $a_{\mu,\lambda} \equiv \frac{\partial a_{\mu}}{\partial \lambda}$ and 
$a_{\mu,\nu\lambda} \equiv \frac{\partial}{\partial \lambda}
\frac{\partial}{\partial z^{\nu}} a_{\mu}$.

If $r$ corresponds to a point $ p \in M$, described in terms of the local
coordinates $ \{ z^{\alpha} \}$, then $ r_{, \mu}(p) \in T_pM$, $ r_{,\mu} (p)=(
\partial y^i/ \partial z^{\mu} ) \partial_i (p)= \partial_{\mu} (p)$ and

\begin{eqnarray}
g_{\mu\nu} (p)& \equiv & \bar{g} (r_{,\mu} , r_{,\nu} ) (p) \nonumber \\
              & =      &ck( - U^{-1} a_{\mu,\lambda} U, -U^{-1}
              a_{\nu,\lambda} U ) (p) =c k(a_{\mu,\lambda}, a_{\nu,\lambda}
              )(p) , \label{eq5.7}
\end{eqnarray}
because $U^{-1}GU$ is an automorphism of $G$ and the Killing form is
invariant by automorphisms. As we see, one of the conditions we need to have
a well defined semi-riemannian submanifold is that the matrix defined by $k(
a_{\mu,\lambda},a_{\nu,\lambda})$ has determinant different of zero. In this
case, one can find $N_1^0$, $N_2^0$, $\ldots$, $N_{\bar{m} -m}^0 \in G$ such that 
\begin{eqnarray}
&& k(a_{\mu,\lambda}, N^0_A) =0 \ \ \hbox{and} \label{eq5.8} \\
&& ck(N^0_A,N^0_B)= \eta_{AB} =\delta_{AB} \epsilon^{\perp}_B, \label{eq5.9}
\end{eqnarray}
where $\epsilon^{\perp}_B= \pm 1$ and the index is denoted by
$\nu^{\perp}$. Note that $ \{ \eta_{AB} \} $, $ \{ \epsilon^{\perp}_B \}$ and $
\nu^{\perp}$ depend on the sign of $c$. That is, if a set $ \{ N^0_A \}$
satisfy (\ref{eq5.8}) and (\ref{eq5.9}) when $c >0$, then the same set will
satisfy (\ref{eq5.8}) and (\ref{eq5.9}) when $c <0$, where now $ \{ \eta_{AB} \}
\rightarrow \{ -  \eta_{AB} \} $, $ \{ \epsilon^{\perp}_B \} \rightarrow 
\{ -\epsilon^{\perp}_B \}$ and $\nu^{\perp} \rightarrow  (\bar{m} -m-\nu^{\perp})$. For
instance, in the example given in the next section $(\bar{m} -m) =1$ and
$\eta_{11}= c/ (\mid c \mid)$.

Define 
\begin{equation}
N_A = U^{-1} N_A^0 U. \label{eq5.10}
\end{equation}
Then $ck(N_A,N_B) = \eta_{AB}$ and, using (\ref{eq5.4}), $N_{A,\mu} =U^{-1}
(N^0_{A,\mu} + [a_{\mu},N^0_A])U$. Note that $\bar{g}(r_{,\mu},N_A)=0$.

Let us employ a natural coordinate system on
$\R^{\bar{m}}_{\bar{\nu}(c)}$. Then $\bar{\Gamma}^k_{ij} =0$ and the GW
equations (\ref{eq4.3}) and (\ref{eq4.4}) can be written as:
\begin{eqnarray}
&& r_{,\alpha\beta} = \Gamma^{\gamma}_{\alpha\beta} r_{,\gamma} 
+ \eta^{AB} b_{A\alpha\beta} N_B \ \ \hbox{and} \label{eq5.11} \\
&& N_{A,\alpha} = \mu^C_{ \, \, \, \, A\alpha} N_C - b_{A\alpha\gamma} g^{\gamma \rho}
r_{,\rho}. \label{eq5.12}
\end{eqnarray}

From (\ref{eq5.11}), $\bar{g}(r_{,\alpha\beta},r_{\rho}) =
  \Gamma^{\gamma}_{\alpha\beta}  g_{\gamma\rho}$. Then $
\Gamma^{\mu}_{\alpha\beta} = g^{\rho\mu} \bar{g} (r_{,\alpha\beta},
  r_{\rho})$. Thus
\begin{equation}
\Gamma^{\mu}_{\alpha\beta}=
cg^{\rho\mu}k(a_{\rho,\lambda},a_{\alpha,\beta\lambda}
-[a_{\alpha,\lambda},a_{\beta}]), \label{eq5.13}
\end{equation}
where $( g^{\rho\mu} )$ is the inverse of the matrix defined by
(\ref{eq5.7}). 

Similarly, from (\ref{eq5.11}),
\begin{equation}
b_{C\alpha\beta}=ck(N^0_C,
[a_{\alpha,\lambda},a_{\beta}]-a_{\alpha,\beta\lambda}) \label{eq5.14}
\end{equation}
and, from (\ref{eq5.12}),
\begin{equation}
b_{C\alpha\beta}=ck(a_{\beta,\lambda}, N^0_{C,\alpha}
+[a_{\alpha},N^0_C]). \label{eq5.15}
\end{equation}
One can verify, using (\ref{eq5.1}), (\ref{eq5.3}) and (\ref{eq5.8}), that
  the expressions (\ref{eq5.14}) and \\ (\ref{eq5.15}) coincide.

From (\ref{eq5.12}),
\begin{equation}
\mu^D_{ \, \, \, \, A\alpha} =c \eta^{BD} k (N^0_B,N^0_{A,\alpha}
+[a_{\alpha},N^0_A]). \label{eq5.16}
\end{equation}
It follows that 
\begin{equation}
\mu_{BA\alpha} =c  k (N^0_B,N^0_{A,\alpha}
+[a_{\alpha},N^0_A]). \label{eq5.17}
\end{equation}

It is possible verify, using (\ref{eq5.3}), that in the expressions 
(\ref{eq5.14}), (\ref{eq5.15}) and \\ (\ref{eq5.17}) 
  $b_{C\alpha\beta} = b_{C\beta\alpha}$ and
$\mu_{DA\alpha} =-\mu_{AD\alpha}$. We show in the appendix that
$\Gamma^{\mu}_{\alpha\beta}$ from the expression (\ref{eq5.13}) and 
$\Gamma^{\mu}_{\alpha\beta}$ obtained from $g_{\mu\nu}$ by (\ref{eq3.1}) are
identical by the use of (\ref{eq5.3}). As in the expression  (\ref{eq3.1}) 
$\Gamma^{\mu}_{\alpha\beta} = \Gamma^{\mu}_{\beta\alpha} $, the same holds in
the expression \\ (\ref{eq5.13}). This can be seen also directly in
(\ref{eq5.13}), because \\ $ a_{\alpha,\beta\lambda} -
[a_{\alpha,\lambda},a_{\beta}] = (1/2) (  a_{\alpha,\beta\lambda} -
[a_{\alpha,\lambda},a_{\beta}] ) + (1/2)  (  a_{\beta,\alpha\lambda} -
[a_{\beta,\lambda},a_{\alpha}] )$.

We are now going to discuss the conditions that must be imposed in order such
construction defines a semi-riemannian submanifold. Let 
$ \phi : \R^m_{\nu} \rightarrow \R^{\bar{m}}_{\bar{\nu} (c)}$ be defined by
$ \phi (z^1, \ldots, z^m) = (y^1, \ldots, y^{\bar{m}})$, where $ \{ y^i \}
$ are the functions defined by (\ref{eq5.5}), evaluated at a fixed value of $\lambda$.
The conditions are: There is an open set $V \subseteq \R^m_{\nu}$ such that 
\begin{itemize}
\item[(a)] $\phi$ is a $C^{\infty}$ differentiable function.
\item[(b)] $a_{1,\lambda}(z^1, \ldots, z^m, \lambda)$, $\ldots$, 
$a_{m,\lambda}(z^1, \ldots, z^m, \lambda)$ are linearly independent elements of $G$.
\item[(c)] $ \det ( k(a_{\mu,\lambda},a_{\nu,\lambda})) \neq 0$.
\end{itemize}

To understand the meaning of the second condition, suppose there is $ p \in V$
such that $r_{,1}(p)$, $\ldots$, $r_{,m}(p)$ are linearly dependent. Then
there are $c_1, \ldots, c_m \in \R$ such that there is at least of them that
satisfies $c_j \neq 0$ and, by (\ref{eq5.6}), \\ $U^{-1}(c_1a_{1,\lambda}+ \cdots
+ c_ma_{m,\lambda})U(p)=0$. Thus $(c_1a_{1,\lambda}+ \cdots
+ c_ma_{m,\lambda})(p)=0$, which would correspond to linearly dependent
$a_{1,\lambda}(p)$, $\ldots$, $a_{m,\lambda}(p)$. Then $r_{,1}(p)$, $\ldots$,
$r_{,m}(p)$ are linearly independent elements of $G$, $\forall p \in V$. It
follows that the $ \bar{m} \times m$ matrix $ \partial(y^1, \ldots,
y^{\bar{m}}) / \partial(z^1, \ldots, z^m)$ corresponding to the differential
of $\phi$ has the maximal rank $m$. Thus $d\phi$ is injective. We have the
conclusion by the first and second conditions that 
$\phi:V \rightarrow \R^{\bar{m}}_{\bar{\nu} (c)}$ is an immersion. As
explained in the previous section, $\forall p \in V \subseteq \R^m_{\nu}$
there is an open set $W$, $p \in W \subseteq V$, such that 
$\phi:W \rightarrow   \R^{\bar{m}}_{\bar{\nu} (c)}$ is an imbedding and
$\phi(W)$ is a submanifold of $ \R^{\bar{m}}_{\bar{\nu} (c)}$.

The last condition implies that the map given by (\ref{eq5.7}) is
nondegenerate. We can take $W$ as a conex open set. Then, as $\phi:W
\rightarrow \phi(W)$ is a homeomorphism, $\phi(W)$ is conex. It is known \cite{Neill} that
if a map $g: \xdif(M) \times \xdif(M) \rightarrow \cal F(M)$ on a conex
manifold $M$ is $\cal F(M)$-bilinear, symmetric and nondegenerate, then it has
constant index.
Thus the last condition implies $\phi(W)$ is a
semi-riemannian submanifold. 

Note that, as we are taking derivatives with respect to the parameter
$\lambda$ in the expressions written in this section, we are implicitly
supposing that differentiability requirements about this parameter are
satisfied.

The field equations are invariant by a gauge transformation, as explained in
the section 2. It is possible to show that the fundamental forms given by
(\ref{eq5.7}), \\ (\ref{eq5.14}) and (\ref{eq5.16}) are invariant by a gauge
transformation given by a $\lambda$-independent group element $g$. That is, in
any matrix representation, $ \partial_{\lambda} g=0$. To do that, as in the
equations (\ref{eq5.8}) and (\ref{eq5.9}), we need to find $ \{ N^{0'}_{A} \}$
such that
\begin{eqnarray*}
&&k(a^g_{\mu,\lambda},N^{0'}_A)=0 \ \ \hbox{and} \\
&&ck(N^{0'}_A,N^{0'}_B)=\eta_{AB},
\end{eqnarray*}
where 
\begin{eqnarray}
&&a^g_{\mu} = ga_{\mu}g^{-1}-\partial_{\mu}g g^{-1} \ \ \hbox{and} \label{eq5.18} \\
&&a^g_{\mu,\lambda} = ga_{\mu,\lambda}g^{-1}.
\end{eqnarray}
We take
\begin{equation}
N^{0'}_A \equiv g N^0_Ag^{-1}. \label{eq5.19}
\end{equation}
Then one can verify that (\ref{eq5.7}), (\ref{eq5.14}) and (\ref{eq5.16})
calculated at $ \{ \{ a_{\mu} \}, \{ N_A^0 \} \}$ and  $ \{ \{ a_{\mu}^g \},
\{ N_A^{0'} \} \}$ are identical.

By last, we emphasize that we are associating to a field theory not just one
but a set of semi-riemannian submanifolds. Different solutions of the field
equations correspond, in general, to different explicit expressions of the
gauge potentials and these last ones correspond, in general, to different submanifolds.

\section{Submanifolds associated to Toda theories}

\setcounter{equation}{0}
\def\theequation{\thesection. \arabic{equation}}

In this section we apply the method of the previous section to Toda theories
associated to semisimple real Lie algebras of finite dimension. In this case,
$ \R^m_{\nu}$ is $\R^2_1 \simeq Mk^2$ and we denote $z^1 \equiv  z$, $z^2 \equiv
\bar{z}$. The gauge potentials are given by (\ref{eq2.3}) and
(\ref{eq2.4}). Then
\begin{eqnarray*}
&&a_{1,\lambda}=-e^{-\lambda}B\varepsilon^-B^{-1} \ \ \hbox{and} \\
&&a_{2,\lambda}=-e^{\lambda}\varepsilon^+.
\end{eqnarray*}

The condition (b) of section 5 is that $a_{1,\lambda}$ and $a_{2,\lambda}$ are
linearly independent. Note that $a_{1,\lambda} \in G_{-1}$ and $a_{2,\lambda}
\in G_1$. Then, if $\varepsilon^+ \neq 0$ and $\varepsilon^- \neq0$, that
condition is satisfied.

Given a semisimple real Lie algebra $G$, suppose there is a grading operator
$Q$ and let $T_n \in G_n$, $T_m \in G_m$. Then, using (\ref{eq5.1}),
$k([Q,T_n],T_m)=-k(T_n,[Q,T_m])$. We see that $k([Q,T_n],T_m)=nk(T_n,T_m)$ and
$k(T_n,[Q,T_m])=mk(T_n,T_m)$. Then $(n+m)k(T_n,T_m)=0$. Thus, if $ (n+m) \neq
0$, then $k(T_n,T_m)=0$.

Let us analize the condition (c) of section 5. We have
\begin{eqnarray*}
&&k(a_{1,\lambda},a_{1,\lambda})=k(a_{2,\lambda},a_{2,\lambda})=0, \\
&&k(a_{1,\lambda},a_{2,\lambda})=k(a_{2,\lambda},a_{1,\lambda})=k(B\varepsilon^-B^{-1},\varepsilon^+).
\end{eqnarray*}
As the Killing form is invariant by authomorphisms, 
\\
$k(B\varepsilon^-B^{-1},\varepsilon^+)=k(\varepsilon^-,B^{-1}\varepsilon^+B)$.
Thus the condition (c) of section 5 is
\begin{equation}
k(B\varepsilon^-B^{-1},\varepsilon^+)=k(\varepsilon^-,B^{-1}\varepsilon^+B)
\neq 0. \label{eq6.1}
\end{equation}

The metric tensor has components given by:
\begin{eqnarray*}
&&g_{11}=g_{22}=0 \ \ \hbox{and} \\
&&g_{12}=g_{21}=ck(B\varepsilon^-B^{-1},\varepsilon^+)=ck(\varepsilon^-,B^{-1}\varepsilon^+B),
\end{eqnarray*}
where $c \neq 0$.

Given $p \in M$, where $M$ denotes the submanifold, by (\ref{eq5.6}),
\[
c_1[U^{-1}(e^{-\lambda}B\varepsilon^-B^{-1})U](p)+c_2[U^{-1}(e^{\lambda}\varepsilon^+)U](p)
\in T_p(M),
\]
$\forall c_1, c_2 \in \R$. Let
\begin{eqnarray}
&&V_1= e^{-\lambda}U^{-1}B\varepsilon^-B^{-1}U +
\frac{e^{\lambda}U^{-1}\varepsilon^+U}{2ck(\varepsilon^+,B\varepsilon^-B^{-1})}
  \ \ \hbox{and} \label{eq6.2} \\
&&V_2= e^{-\lambda}U^{-1}B\varepsilon^-B^{-1}U -
\frac{e^{\lambda}U^{-1}\varepsilon^+U}{2ck(\varepsilon^+,B\varepsilon^-B^{-1})}.
\label{eq6.3}
\end{eqnarray}
Then $V_1(p)$ and $V_2(p) \in T_p(M)$, $\forall p \in M$ and
$g(V_1,V_1)=ck(V_1,V_1)=1$, \\
$g(V_2,V_2)=-1$ and $g(V_1,V_2)=0$ (Note that
$(V_1,V_2)\mid_{c>0}=(V_2,V_1)\mid_{c<0}$). Thus the index associated to the
submanifold is $\nu^{sub}=1$ and does not depend on the sign of $c$. We have 
$\nu^{sub} \mid_{c>0}+\nu^{sub}\mid_{c<0}=m=2$. Then $(\nu^{sub} \mid_{c>0}=1) 
\Longleftrightarrow (\nu^{sub} \mid_{c<0}=1)$. Note that $\nu^{sub}=\nu$, where
$\nu=1$ is the index associated to $\R^2_1 \simeq Mk^2$.

The inverse of $g$ has components given by:
\begin{eqnarray*}
&&g^{11}=g^{22}=0 \ \ \hbox{and} \\
&&g^{12}=g^{21}=\frac{1}{ck(B\varepsilon^-B^{-1},\varepsilon^+)}=
\frac{1}{ck(\varepsilon^-,B^{-1}\varepsilon^+B)}.
\end{eqnarray*}

The Christoffel symbols can be obtained using(\ref{eq2.3}), (\ref{eq2.4}),
(\ref{eq5.13}) and the fact that if $(n+m) \neq 0$ then $k(G_n,G_m)=0$:
\begin{eqnarray*}
&&\Gamma^1_{11}=\frac{k(\varepsilon^+,\partial_1(B\varepsilon^-B^{-1}))}
{k(\varepsilon^+,B\varepsilon^-B^{-1})}, \\
&&\Gamma^1_{12}=\Gamma^1_{21}=
\frac{k(\varepsilon^+,\partial_2(B\varepsilon^-B^{-1})
 +[B\varepsilon^-B^{-1},\partial_2BB^{-1}])}
{k(\varepsilon^+,B\varepsilon^-B^{-1})}, \\
&&\Gamma^2_{11}=\Gamma^2_{12}=\Gamma^2_{21}=\Gamma^1_{22}=0 \ \ \hbox{and} \\
&&\Gamma^2_{22}=\frac{k([B\varepsilon^-B^{-1},\varepsilon^+],\partial_2BB^{-1})}
{k(\varepsilon^+,B\varepsilon^-B^{-1})}.
\end{eqnarray*}

The gaussian curvature $K=-R_{1212}=-g_{1\mu}R^{\mu}_{212}$, can be obtained using
(\ref{eq3.2}) and the metric tensor components and Christoffel symbols given
in this section:
\[
K=ck(B\varepsilon^-B^{-1},\varepsilon^+)\partial_1
\left[ \frac{k([B\varepsilon^-B^{-1},\varepsilon^+],\partial_2BB^{-1})}
{k(\varepsilon^+,B\varepsilon^-B^{-1})} \right].
\]
As we explained in the section 3, the Riemann tensor, the Ricci tensor and the
scalar curvature of the submanifold are given in terms of the gaussian curvature.

We now give an example of an abelian Toda theory. Let $G$ be the simple real Lie algebra $sl(2,\R)$,
which is a non-compact real form of the simple complex Lie algebra $A_1$. The
commutation relations are: $[H,E_{ \pm \alpha } ]= \pm \alpha E_{ \pm \alpha }$ and 
$[E_{\alpha},E_{-\alpha}]=2 \alpha H / \alpha^2$. The Killing form is
$k(H,H)=1$, $k(E_{\alpha},E_{-\alpha})=2/ \alpha^2$ and zero in the other cases.
In the Chevalley basis $ \{ h, E_{\alpha}, E_{-\alpha} \}$, we have:
$h \equiv 2\alpha H / \alpha^2$, $[h,E_{ \pm \alpha}] =\pm 2 E_{ \pm \alpha}$, 
$[E_{\alpha},E_{-\alpha}]=h$ and $k(h,h)=4 / \alpha^2$ \cite{Cornwell}.

Note that
\begin{eqnarray*}
&&k \left( \frac{\alpha}{2} (E_{\alpha}+E_{-\alpha}),\frac{\alpha}{2}
(E_{\alpha}+E_{-\alpha}) \right) =1, \\ 
&&k \left( \frac{\alpha}{2} (E_{\alpha}-E_{-\alpha}),\frac{\alpha}{2}
(E_{\alpha}-E_{-\alpha}) \right) =-1 \ \ \hbox{and} \\ 
&&k \left( \frac{\alpha}{2} (E_{\alpha}+E_{-\alpha}),\frac{\alpha}{2}
(E_{\alpha}-E_{-\alpha}) \right) =0. 
\end{eqnarray*}
One can verify that $ \{ H, \frac{\alpha}{2} (E_{\alpha}+E_{-\alpha}), 
\frac{\alpha}{2} (E_{\alpha}-E_{-\alpha}) \} $ is an orthonormal basis for
$sl(2,\R)$ with respect to the Killing form. Thus $\bar{m} = \dim(sl(2,\R))=3$
and $\bar{\nu}=1$.

The fundamental weight $\Lambda$ is defined by $2\alpha\Lambda / \alpha^2
=1$. We define $Q \equiv 2 \Lambda H / \alpha^2$. Then $h \in G_0$,
$E_{\alpha} \in G_1$ and $E_{-\alpha} \in G_{-1}$. We define $\varepsilon^+=\mu^+
E_{\alpha}$ and  $\varepsilon^-=\mu^- E_{-\alpha}$, where $\mu^+, \mu^- \in
\R$, $\mu^+ \neq 0$ and $\mu^- \neq 0$.
The group element $B$ is parametized as $B \equiv \exp (\varphi h)$, where $\varphi
\in {\cal F} (Mk^2)$, that is, $\varphi$ is a $C^{\infty}$ differentiable
function $ \varphi : Mk^2 \rightarrow \R$. The gauge potentials are:
\begin{eqnarray}
&&a_1= \mu^- e^{-\lambda} e^{-2\varphi} E_{-\alpha} \ \ \hbox{and} \label{eq6.4}
 \\
&&a_2=- \mu^+ e^\lambda E_{\alpha} - \partial_2 \varphi h. \label{eq6.5}
\end{eqnarray}
The field equation is $\partial_1\partial_2 \varphi = \mu^+\mu^-e^{-2\varphi}$.
Using the $(t,x)$ variables defined in the section 2, $
(\partial_t^2-\partial_x^2)\varphi = 4 \mu^+ \mu^- e^{-2\varphi}$. Note that 
\\ $k(B\varepsilon^-B^{-1},\varepsilon^+)=(2/ \alpha^2) \mu^+ \mu^- e^{-2\varphi}
\neq 0$. Thus the condition(\ref{eq6.1}) is satisfied.

The metric tensor components are given by:
\begin{eqnarray*}
&&g_{11}=g_{22}=0 \ \ \hbox{and} \\
&&g_{12}=g_{21}= \frac{2c}{\alpha^2} \mu^+ \mu^- e^{-2\varphi}.
\end{eqnarray*}

The inverse of $g$ has components given by:
\begin{eqnarray*}
&&g^{11}=g^{22}=0 \ \ \hbox{and} \\
&&g^{12}=g^{21}=\frac{\alpha^2e^{2\varphi}}{2c\mu^+\mu^-}.
\end{eqnarray*}

The Christoffel symbols are given by:
\begin{eqnarray*}
&&\Gamma^1_{11}=-2\partial_1\varphi, \\
&&\Gamma^2_{11}=\Gamma^1_{12}=\Gamma^1_{21}=\Gamma^2_{12}=\Gamma^2_{21}=\Gamma^1_{22}=0 \ \
\hbox{and} \\
&&\Gamma^2_{22}=-2\partial_2\varphi.
\end{eqnarray*}

The gaussian curvature is given by:
\[
K=-\frac{4c}{\alpha^2} (\mu^+\mu^-)^2 e^{-4\varphi},
\]
where $c \neq 0 $.Note that:
\begin{itemize}
\item[(a)] $\varphi=$constant is not a solution of the field equation. Thus
  $K$ is not constant.
\item[(b)]If $c>0$, then $K(p)<0$, $\forall p \in M$. If $c<0$, then $K(p)>0$,
$\forall p \in M$.
\end{itemize}

As shown in the section 5, we need a set of $(\bar{m}-m)$ elements $ \{ N^0_A \}$
satisfying \\ (\ref{eq5.8}) and (\ref{eq5.9}). In our example,
$(\bar{m}-m)=1$. Then $N^0_1=H/ \sqrt{ \mid c \mid }$. As $k(H,H)=1$,
$\eta_{11}=\eta^{11}=c / \mid c \mid$.

From (\ref{eq5.14}), (\ref{eq5.15}), (\ref{eq6.4}) and (\ref{eq6.5}),
\begin{eqnarray*}
&&b_{111}=b_{122}=0 \ \ \hbox{and} \\
&&b_{112}=b_{121}= - \frac{2c \mu^+\mu^- e^{-2\varphi}}{\alpha \sqrt{ \mid c \mid}}.
\end{eqnarray*}

In our example the submanifold is a hypersurface $(\bar{m}=m+1)$. Then, as
shown in the section 4, we have that $\mu^1_{ \, \, \, \, 11}=\mu^1_{ \, \, \,
  \, 12}= 0$.
 
Given $p \in M$, then $(V_1(p),V_2(p))$ (equations (\ref{eq6.2}) and
(\ref{eq6.3})) is an orthonormal basis at $T_p(M)$ $(\epsilon_1=1$ and
$\epsilon_2=-1)$. As $( r_{,1}=\partial_1, r_{,2}=\partial_2)$, they can be written as:
\begin{eqnarray*}
&&V_1= \partial_1 + \frac{1}{2ck(\varepsilon^+,B\varepsilon^-B^{-1})} \, \partial_2
\ \ \hbox{and} \\
&&V_2= \partial_1 - \frac{1}{2ck(\varepsilon^+,B\varepsilon^-B^{-1})} \, \partial_2.
\end{eqnarray*} 
The mean curvature vector field $\vec{\cal H}$ defined in the equation
(\ref{eq4.5}) is given by:
\[
\vec{\cal H} = \frac{1}{2} [ \Pi(V_1,V_1)-\Pi(V_2,V_2) ].
\]
Using that $\Pi$ is $ \cal F (M) $-bilinear,
$\Pi(\partial_{\mu},\partial_{\nu})=\eta^{AB}b_{B\mu\nu}N_A$
 and $N_A=U^{-1}N^0_AU$:
\[
\vec{\cal H}= - \frac{\alpha}{c} U^{-1}HU=-\frac{\alpha^2}{2c} U^{-1}hU.
\]
Note that:
\begin{itemize}
\item[(a)] $\bar{g}( \vec{\cal H} , \vec{\cal H} )(p)= ck( \vec{\cal H} ,
  \vec{\cal H} )(p)= \alpha^2 / c =$ constant, $\forall p \in M$. The
  vector field $\vec{\cal H}$ has the same causal character at all the points
  of the submanifold. That is, if $ c>0 \, (c<0)$, then $\vec{\cal H} (p)$ is a
  spacelike (timelike) vector, $\forall p \in M$.
\item[(b)] $\vec{\cal H} = - \alpha  \sqrt{ \mid c \mid } N_1 / c.$ Then
\[
D^{\perp}_V \vec{\cal H}= - V^{\beta} \partial_{\beta} \left( \alpha 
\frac{ \sqrt{ \mid c \mid }}{c} \right) N_1 - \alpha \frac{ \sqrt{\mid c \mid}}{c} D^{\perp}_V N_1 =0,
\]
$\forall V \in \xdif (M)$, because $ ( \alpha \sqrt{ \mid c \mid } / c ) $ is
constant,  $\mu^1_{ \, \, \, \, 11}=\mu^1_{ \, \, \,
  \, 12}= 0$ and \\ $ D^{\perp}_V N_1= V^{\beta} \mu^1_{ \, \, \, \, 1\beta}N_1=0$.
\end{itemize}

By last, we want to analyze the structure of the set $ \{ N_A^0 \} $ in
the case of
an abelian Toda theory associated to a higher rank algebra. Let $G$ be the
simple real Lie algebra $sl(3, \R)$, which is a non-compact real form of the
simple complex Lie algebra $A_2$. In an analogous way to that described in the
case of $sl(2, \R)$, the commutation relations of $sl(3, \R)$ associated to
the basis  $ \{ \{ H_i \} , \{ E_{\alpha} \} \} $  or   
$ \{ \{ h_i \} , \{ E_{\alpha} \} \} $ can be obtained from the commutation
relations of $A_2$ \cite{Cornwell}. We define
\[
Q \equiv 2 \frac{\Lambda_1 \cdot H}{\alpha_1^2} + 
         2 \frac{\Lambda_2 \cdot H}{\alpha_2^2},
\]
where $\Lambda_1, \, \Lambda_2$ are the fundamental weights. Then 
$ \{ h_1,h_2 \} \in G_0 $, $ \{ E_{\alpha_1}, E_{\alpha_2} \} \in G_1$, 
$ \{ E_{-\alpha_1}, E_{-\alpha_2} \} \in G_{-1}$, 
$ \{ E_{\alpha_1+\alpha_2} \} \in G_2 $ and $ \{ E_{-\alpha_1-\alpha_2} \}
\in G_{-2} $, where $ h_i = 2 \alpha_i \cdot H / \alpha_i^2$.

We define $\varepsilon^+ = \mu^+(E_{\alpha_1}+E_{\alpha_2})$ and 
          $\varepsilon^- = \mu^-(E_{-\alpha_1}+E_{-\alpha_2})$, where $\mu^+,
          \mu^- \in \R$, $\mu^+ \neq 0$ and $\mu^- \neq 0$.
The group element is parametrized as $B \equiv \exp
(\varphi_1h_1+\varphi_2h_2)$, where $\varphi_1, \varphi_2 \in {\cal F}(Mk^2)$,
$\varphi_1: Mk^2 \rightarrow \R$, $\varphi_2:Mk^2 \rightarrow \R$. Note that
\\ $k(B\varepsilon^-B^{-1},\varepsilon^+)=
\mu^+\mu^-(e^{-2\varphi_1+\varphi_2}+e^{\varphi_1-2\varphi_2}) \neq 0$, by the
normalization $\alpha^2 =2$ for all the roots. Thus the condition
(\ref{eq6.1}) is satisfied.

The gauge potentials are given by:
\begin{eqnarray*}
&&a_1=
e^{-\lambda} \mu^-(e^{-2\varphi_1+\varphi_2}E_{-\alpha_1}+e^{\varphi_1-2\varphi_2}E_{-\alpha_2})
  \ \ \hbox{and} \\
&&a_2= - e^{\lambda} \mu^+(E_{\alpha_1}+E_{\alpha_2})-
( \partial_2 \varphi_1h_1+\partial_2\varphi_2h_2).
\end{eqnarray*}

We have $( \bar{m}-m) =8-2=6$ elements in the set $ \{ N_A^0 \} $:
\begin{eqnarray*}
&&N_1^0= \frac{H_1}{\sqrt{\mid c \mid}}, \, N_2^0= \frac{H_2}{\sqrt{\mid c \mid}},
\, N_3^0= \frac{1}{\sqrt{ 2 \mid c
    \mid}}(c_1E_{\alpha_1}-c_2E_{\alpha_2}-E_{-\alpha_1}+E_{-\alpha_2}), \\
&&N_4^0= \frac{1}{\sqrt{ 2 \mid c
    \mid}}(c_1E_{\alpha_1}-c_2E_{\alpha_2}+E_{-\alpha_1}-E_{-\alpha_2}), \\ 
&&N_5^0= \frac{1}{ \sqrt{2 \mid c \mid } } (E_{\alpha_1+\alpha_2}+E_{-\alpha_1-\alpha_2}), \,
N_6^0= \frac{1}{ \sqrt{2 \mid c \mid } }
(E_{\alpha_1+\alpha_2}-E_{-\alpha_1-\alpha_2}), \,
\end{eqnarray*}
where
\[
c_1= \frac{ \exp
  [(3/2)(\varphi_1-\varphi_2)]}{2\cosh[(3/2)(\varphi_1-\varphi_2)]} \ \
  \hbox{and} \ \ 
c_2= \frac{ \exp
  [-(3/2)(\varphi_1-\varphi_2)]}{2\cosh[(3/2)(\varphi_1-\varphi_2)]}.
\]

Note that:
\begin{itemize}
\item[(a)] Each one of the elements in the set $ \{ N^0_A \} $ belongs to one
  of the subspaces: $G_0$, $G_1 \oplus G_{-1}$, $G_2 \oplus G_{-2}$.
\item[(b)] The elements associated to the subspaces $G_0$ and  
$G_2 \oplus G_{-2}$ do not depend on the variables $z_1, z_2$. That is, they
do not depend on the field variables.
\end{itemize}
In a similar way, one can construct the set  $ \{ N^0_A \} $ in the case of a
general Toda theory.

\section*{Appendix}

We want to show that
\[
\Gamma^{\mu}_{\alpha\beta}=\frac{1}{2}cg^{\rho\mu}[\partial_{\alpha}k(a_{\beta,\lambda},a_{\rho,\lambda})
+\partial_{\beta}k(a_{\alpha,\lambda},a_{\rho,\lambda})-\partial_{\rho}k(a_{\alpha,\lambda},a_{\beta,\lambda})]
\]
is identical to $\Gamma^{\mu}_{\alpha\beta}$ given by the equation
(\ref{eq5.13}). Taking the derivatives in the order they appear, we have six terms:
 $\Gamma^{\mu}_{\alpha\beta}=(\Gamma^{\mu}_{\alpha\beta})^1+ \cdots
 +(\Gamma^{\mu}_{\alpha\beta})^6$. For instance, using \\ (\ref{eq5.1}) and (\ref{eq5.3}),
\begin{eqnarray*}
(\Gamma^{\mu}_{\alpha\beta})^2+(\Gamma^{\mu}_{\alpha\beta})^5 &=&
\frac{1}{2} cg^{\rho\mu}[k(a_{\beta,\lambda},a_{\rho,\alpha\lambda})
-k(a_{\alpha,\rho\lambda},a_{\beta,\lambda})] \\
&=& \frac{1}{2} c g^{\rho\mu}k(a_{\beta,\lambda},a_{\rho,\alpha\lambda}-a_{\rho,\alpha\lambda}
-[a_{\alpha,\lambda},a_{\rho}]-[a_{\alpha},a_{\rho,\lambda}]) \\
&=& \frac{1}{2} c
g^{\rho\mu}[k(a_{\rho,\lambda},[a_{\alpha},a_{\beta,\lambda}])
-k(a_{\rho},[a_{\beta,\lambda},a_{\alpha,\lambda}])].
\end{eqnarray*}

Similarly,
$(\Gamma^{\mu}_{\alpha\beta})^1+(\Gamma^{\mu}_{\alpha\beta})^3=
(1/2) cg^{\rho\mu}
k(a_{\rho,\lambda},2a_{\alpha,\beta\lambda}-[a_{\alpha,\lambda},a_{\beta}]
-[a_{\alpha},a_{\beta,\lambda}])$
\\ and  
$(\Gamma^{\mu}_{\alpha\beta})^4+(\Gamma^{\mu}_{\alpha\beta})^6=
(1/2) cg^{\rho\mu}[k(a_{\rho,\lambda},[a_{\beta},a_{\alpha,\lambda}])
-k(a_{\rho},[a_{\alpha,\lambda},a_{\beta,\lambda}])]$.

Thus
$
\Gamma^{\mu}_{\alpha\beta}=cg^{\rho\mu}k(a_{\rho,\lambda},a_{\alpha,\beta\lambda}-[a_{\alpha,\lambda},a_{\beta}]).
$

\end{document}